\documentclass[preprint2]{emulateapj}

\pdfoutput=1
\usepackage{multirow}
\usepackage{url,natbib}
\newcommand{\approgsim}[2]{\mathrel{\vcenter{
  \offinterlineskip\halign{\hfil$##$\cr
    #1\geq\cr\noalign{\kern2pt}#1\sim\cr\noalign{\kern-2pt}}}}}

\newcommand{\apgs}{\mathrel{\vcenter{
  \offinterlineskip\halign{\hfil$##$\cr
    >\cr\noalign{\kern2pt}\simeq\cr\noalign{\kern-2pt}}}}}

\newcommand{\apgss}{\mathrel{\vcenter{
  \offinterlineskip\halign{\hfil$##$\cr
    >\cr\noalign{\kern2pt}\simeq\cr\noalign{\kern-2pt}}}}}

\slugcomment{Submitted to {\it The Astrophysical Journal}}
\shorttitle{Earth's disk-integrated thermal emission}
\shortauthors{G\'{o}mez-Leal I., Pall\'{e} E. and Selsis F.}

\begin{document}

\title{PHOTOMETRIC VARIABILITY OF THE DISK INTEGRATED INFRARED EMISSION OF THE EARTH}

\author{\scshape I. G\'{o}mez-Leal\altaffilmark{1,}\altaffilmark{2}, E. Pall\'{e}\altaffilmark{3}, and F. Selsis\altaffilmark{1,}\altaffilmark{2}}
\email{gomezleal@obs.u-bordeaux1.fr, epalle@iac.es,}
\email{selsis@obs.u-bordeaux1.fr}
\altaffiltext{1}{Univ. Bordeaux, LAB, UMR 5804, F-33270, Floirac, France.}
\altaffiltext{2}{CNRS, LAB, UMR 5804, F-33270, Floirac, France.}
\altaffiltext{3}{Instituto de Astrof\'{i}sica de Canarias, V\'{i}a L\'{a}ctea s/n, La Laguna E-38205, Tenerife, Spain.}

\begin{abstract}
	Here we present an analysis of the global-integrated mid-infrared emission flux of the Earth based on data derived from satellite measurements. We have studied the photometric annual, seasonal, and rotational variability of the thermal emission of the Earth to determine which properties can be inferred from the point-like signal. We find that the analysis of the time series allows us to determine the 24 hr rotational period of the planet for most observing geometries, due to large warm and cold areas, identified with geographic features, which appear consecutively in the observer's planetary view. However, the effects of global-scale meteorology can effectively mask the rotation for several days at a time. We also find that orbital time series exhibit a seasonal modulation, whose amplitude depends strongly on the latitude of the observer but weakly on its ecliptic longitude. As no systematic difference of brightness temperature is found between the dayside and nightside, the phase variations of the Earth in the infrared range are negligible. Finally, we also conclude that the phase variation of a spatially unresolved Earth--Moon system is dominated by the lunar signal.
\end{abstract}


\section{Introduction}
Finding and characterizing habitable exoplanets is one of the key objectives of 21st century astrophysics. Transit spectroscopy already allows us to constrain some of the atmospheric properties of hot exoplanets as composition \citep{Tinetti2007}, temperature structure (e.g., \cite{Knutson2008}) or circulation \citep{Snellen2010}. The phase variations of the thermal emission or reflected light of hot exoplanets can be extracted from combined star+planet photometry, providing another powerful tool to characterize the climate of exoplanets. This has been achieved in transiting configurations (e.g.,  \cite{Knutson2007}), including one terrestrial-mass planet \citep{Batalha2011}, but also for non-transiting hot Jupiters (e.g., \cite{Cowan2007}). In theory, the thermal orbital light curves of non-transiting short-period rocky planets can also be measured with combined light photometry and used to detect atmospheric species \citep{Selsis2011} or to constrain the radius, albedo and inclination of the planet \citep{Maurin2012}. Transit spectroscopy and orbital light curve measurements may be achievable with the \textit{James Webb Space Telescope} or the \textit{Exoplanet Characterization Observatory} \citep{Tinetti11} for a few habitable planets transiting M stars that can hopefully be found in the solar vicinity \citep{Belu2010, Palle2011}. 
The characterization of an Earth analog around a G star by transit spectroscopy is much more challenging for several reasons: G stars are less numerous than M stars, the transit probability in the habitable zone is 10 times lower than it is for M stars, the orbital period (and thus the duration between two transits) is significantly longer for G stars, and the planet-to-star contrast ratio less favorable for secondary eclipse spectroscopy. 
Direct detection thus seems to be necessary to study Earth analogs around G stars and several instrument concepts have been proposed \citep{Traub2006, Danchi2007, Trauger2007, Cash2009}. Depending on the concept, it is either the light scattered in the visible or the infrared emission that can be detected. Spectroscopy and photometry can then be used to derive some of the planet properties. An important step toward these ambitious programs is to determine what level of characterization could be achieved when observing the Earth as a distant \textit{pale infrared dot}. In the optical range, broadband photometry can potentially give us information about the albedo and the cloud cover. The uneven distribution of oceans and continents enables the measurement of the 24 hr rotation period by the autocorrelation of the signal for several viewing inclination angles despite the presence of active weather \citep{Palle2004, Palle2008}. Visible and near-infrared spectroscopical studies have been made by, e.g., \citet{Hearty2009}, \citet{Cowan2011}, \citet{Robinson2011a} and  \citet{Livengood2011} observing rotational and seasonal variations of the Earth spectrum and their influence on the detectability of the spectral signatures of habitability and life.

In this paper, we provide an integrated mid-infrared (4--50~$\mu$m) photometric time series model of the Earth, with 3 hr time resolution, constructed from over 22 years of available satellite data.  From this geographically resolved data set, we derived the disk-integrated photometric signal of the Earth, seen as a point-source planet. The paper is organized as follows: in Section 2, we describe our model, the input data, and the assumptions that are considered. In Section 3, we describe the analysis of the time series and we discuss the retrieving of the rotation period, the variability of the signal, and the case of the Earth--Moon system. In Section 4, we summarize our conclusions.

\begin{figure*}[th!]
\includegraphics[width=1.0\textwidth]{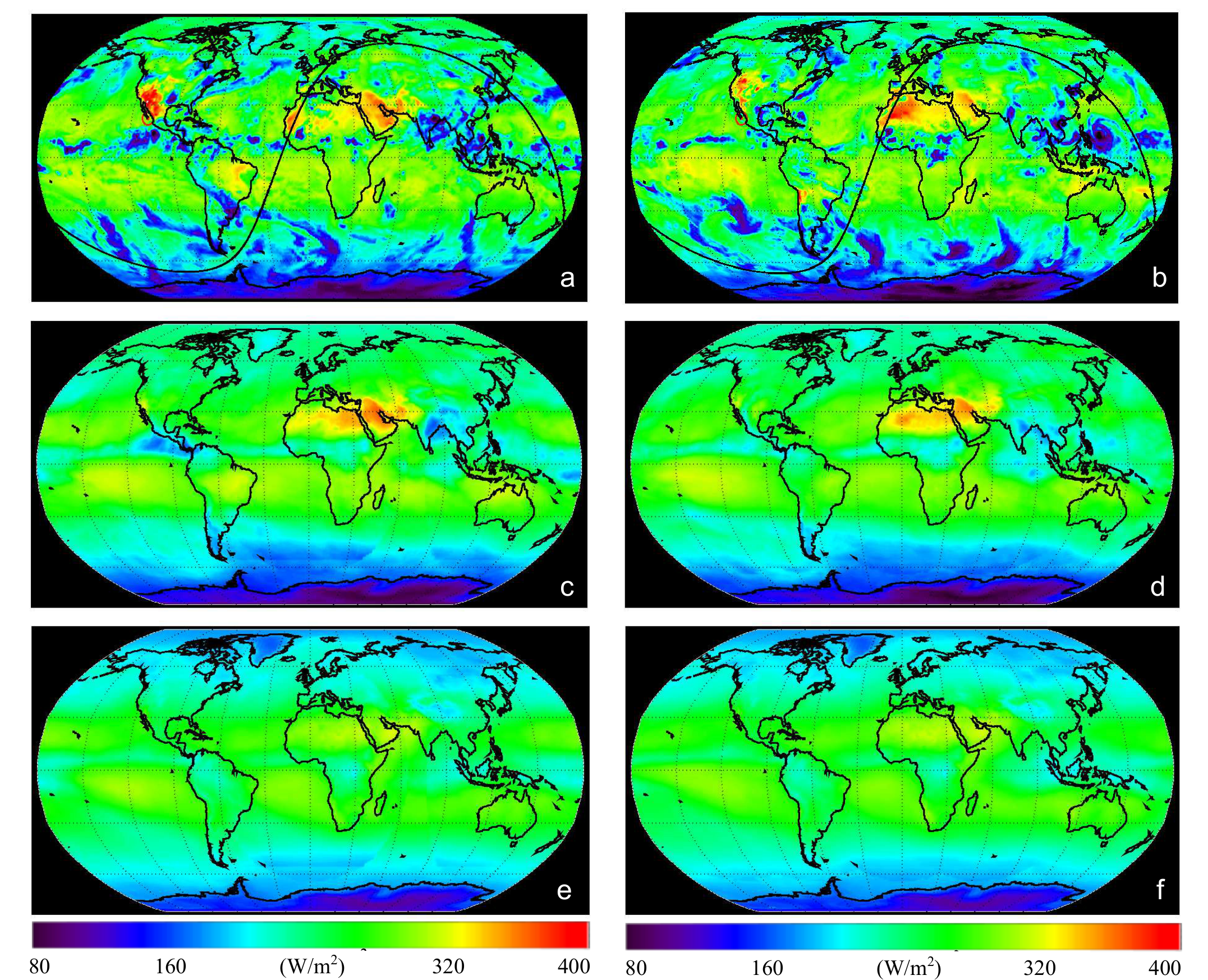}
\caption{Maps of Earth's outgoing mid-infrared radiation. Average over the period 18:00--21:00 UT of 1987 July 1 (a) and 2001 July 1 (b) together with the subsolar point and the terminator at the mean time. At that time the brighter regions of the Earth correspond to the deserts of Sahara, Arabian Peninsula, Atacama, and Arizona. Average over the months 1987 July (c) and 2001 July (d). Average over the years 1987 (e) and 2001 (f).}
\label{fig1}
\end{figure*}

\section{Methods}

The Sun outputs most of its energy in the visible and near-infrared range of the electromagnetic spectrum which is absorbed by the Earth and re-emitted to space in the mid-infrared range. The equilibrium temperature of the planet can be generally defined as:

\begin{equation}
T_{\rm{eq}}={ \left[{ \frac{F_{\rm{S}}(1-A)}{4 \sigma }}\right]}^{1/4} = {\left[{ \frac{L_{\rm{S}}(1-A)}{16 \pi a^{2} \sigma}}\right]}^{1/4},
\end{equation}

where $F_{\rm{S}}$ is the solar incident flux, $A$ is the terrestrial albedo, $\sigma$ is the Stefan--Boltzmann constant,  $L_{\rm{S}}$ is the solar luminosity and $a$ is the Earth--Sun distance. For the Earth, this temperature is approximately \textit{T$_{\rm{eq}}$}$\thicksim$255 K and the emission reaches its maximum at 10 $\micron$ although its spectral distribution is strongly affected by the broad absorption bands of atmospheric greenhouse gases. The emission that a remote observer would detect, however, will be highly dependent on the season, the planetary region that is viewed and the cloud distribution at that time. Here, we derive the variability and periodicities of this integrated flux as it would be seen from a remote perspective, using satellite data and a geometrical model to build the point-like signal received by the observer. The interest of this work lies in the facts that land masses can emit more infrared radiation than ocean areas, these latter have a bigger thermal inertia that make them more resistant to temperature changes and the atmosphere of the Earth is optically thin enough in some windows of the thermal infrared to observe the influence of surface features on the emission at the top of the atmosphere (TOA).

\subsection{Data}
To construct the time series of the emitted infrared flux, we have used TOA all-sky upward longwave (LW) flux integrated over the 4-50 $\mu$m wavelength interval. These data were obtained from the NASA Langley Research Center Atmospheric Sciences Data Center, they are part of the NASA/GEWEX SRB Project\footnote{http://www.gewex.org/} \citep{Suttles1989, Gupta1992} and they cover a 22 year period from 1983 to 2005. The GEWEX LW algorithm used to create these data \citep{Fu1997} uses a thermal infrared radiative transfer code with input parameters derived from the International Satellite Cloud Climatology Project\footnote{http://isccp.giss.nasa.gov/} (ISCCP; \cite{Rossow1996}, the Goddard EOS Data Assimilation System level-4\footnote{http://daac.gsfc.nasa.gov/} (GEOS-4), the Total Ozone Mapping Spectrometer\footnote{http://jwocky.gsfc.nasa.gov/} (TOMS) archive, and the TIROS Operational Vertical Sounder\footnote{http://www.ozonelayer.noaa.gov/action/tovs.htm} (TOVS) data set. The data have a time resolution of 3 hr over the whole globe, and a spatial resolution of $1^{\circ}$x$1^{\circ}$ square cells in latitude and longitude. Typical maps of the outgoing mid-infrared radiation of the Earth, directly represented from the GEWEX data, are shown in Figures~\ref{fig1}(a) and (b). The maps represent the average flux over the period 18:00--21:00 UT for 1987 July 1 and 2001 July 1, respectively. Some desert regions are noticeable, such as Sahara, Arizona or Atacama. Some cold and humid regions are remarkable too, such as Indonesian clouds, or Antarctica.

\begin{figure}[th!]
\begin{center}
\includegraphics[width=1.\columnwidth]{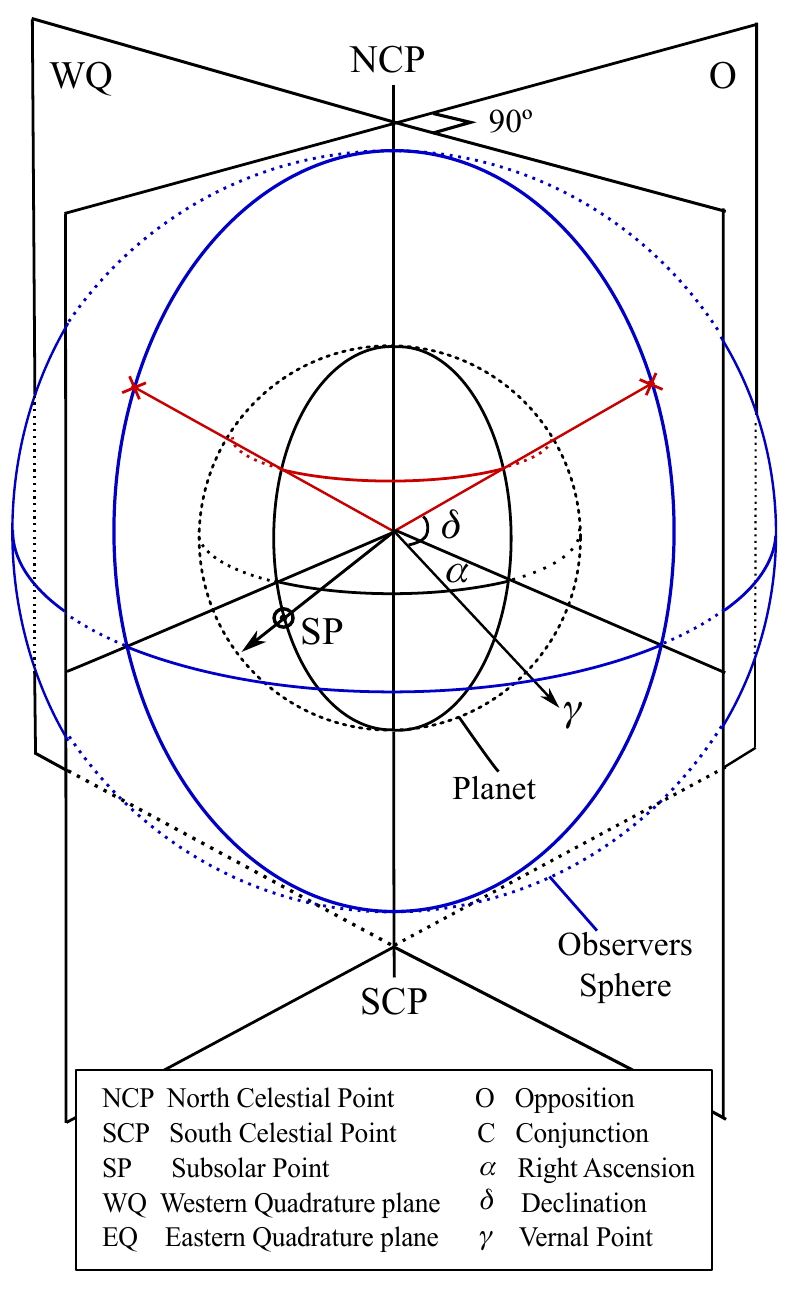}
\caption{Diagram of the model. We take into account the four observers per declination angle at a distance of 10 pc in the planes defined by the directions of opposition (O), western quadrature (WQ), conjunction(C), and eastern quadrature (EQ) and the rotational axis. The intersection of the direction vectors (red) with the sphere of 10 pc (blue) define their position. The sub-observer point is defined by the intersection with the planetary surface (black).}
\label{fig2}
\end{center}
\end{figure}

\subsection{Geometrical Model}
In order to simulate the infrared observations of the Earth seen from any given direction, we have built a geometrical model of the emitted radiation of the Earth. This model calculates at any time the fraction of the planetary disk exposed in the direction of the remote observer and computes the integrated flux F$_{\rm{obs}}$ F$_{{obs}}$received by the observer at a distance $d$:

\begin{equation}\label{eqF}
F_{\rm{obs}}(t)=\frac{1}{d^{2}}\sum_{i}\left [ a_{i} \cdot \rm{cos}\theta_{i}\cdot \left ( \frac{F_{i}(t)}{\pi } \right ) \right ],
\end{equation}

where $a_i$ is the area of each grid cell, $\rm{cos}\theta_{i}$ is the cosine of the angle between the surface normal and the observer's line of sight, $F_{i}$ is the TOA flux given by GEWEX ( with a resolution of 3 hr) and $F_{i}$/$\pi$ is the specific intensity in the Lambertian approximation. We obtain a point-like source emission every 3 hr.

 With the purpose of illustrating several possible planet viewing geometries, we have defined our observer position by equatorial coordinates at a radius distance of 10 pc. The equivalent of the declination angle on the planet surface gives us the latitude of the sub-observer point, which remains constant along the orbit. The position of the planet in the orbit and the position of the sub-observer's point with time are derived from the ephemerides of the substellar point. The ephemeris data are taken from the JPL Horizons System\footnote{http://ssd.jpl.nasa.gov/?horizons} \citep{Gio96}. 

In order to understand and interpret the variation of the Earth emission in relation with geographic features and climates, it is relevant to locate the sub-observer point by its geographic coordinates. Indeed, its latitude remains constant with time because precession and nutation are negligible during an orbit, and its longitude simply varies as $\omega t$, $\omega$ being the rotation rate of the Earth (with the exceptions of the polar cases where the planetary geometry does not change with time). Therefore, we can define the sub-observer point by its latitude and by its longitude at a given time. This time is set to January 1, 0:00 UT of the year considered. We consider four different initial longitudes for the sub-observer point: the meridian of the subsolar point, the morning terminator, the meridian of the antisolar point and the evening terminator. We call these initial observing geometries conjunction (C), western quadrature (WQ), opposition (O) and eastern quadrature (EQ), respectively, although these terms should normally apply only to an observer located in the ecliptic plane (Figure~\ref{fig2}).

It is important to note two facts: the first is that each observer sees the planet at a certain local hour during a whole rotation period and the second, that observers that are placed at the same latitude but different longitudes see the same region of the planet at different local hours (if the (O) observer sees a given region of the planet during the winter midnight and the summer noon, the (C) observer sees the same region during the winter noon and the summer midnight). Furthermore, an observer over a polar latitude sees the same hemisphere of the planet along the time of observation. In the need of a reference for time, we define its ``local hour'' as the Universal Time (UT). In this case, the seasonal variability and the diurnal variability (daily change in temperature in a certain region) produce the variation of the signal and not the changing of the planetary view (rotational variability), as we see in Section 3.2.

\subsection{Limb Darkening}
In order to convert the TOA flux into the disk-integrated flux received by a distant observer, we assume an isotropic (Lambertian) distribution of specific intensities at the TOA. In reality, specific intensities are not isotropic, producing limb-darkening or -brightening depending on the thermal structure and composition of the atmosphere, as well as the wavelength. This approximation gives an exact result only if (1) the atmosphere is uniform (its structure and composition is the same at all latitudes and longitudes) and (2) the TOA flux has been obtained by integrating over all angles the actual distribution of specific intensities. If these conditions are fulfilled, then the angular integration over the planetary disk is equivalent to the angular integration of the outgoing specific intensities done to compute the local flux at the TOA. Although condition (1) is obviously not fulfilled, the geographic variations of the atmospheric structure and the composition of Earth's atmosphere are not steep and, as a consequence, significant departures are expected only for peculiar regions (for instance, the Sahara) when observed at the edge of the planetary disk. However, in the case where these regions do not remain all the time too far ($\gtrsim$70$\degr$) from the sub-observer point, this effect should not affect the amplitude nor the periodicity of the photometric variations but only their shape. In order to evaluate the error caused (for a uniform atmosphere) when the outgoing flux is inferred from one single direction of propagation (two-streams approximation), we have used the Phoenix\footnote{http://www.hs.uni-hamburg.de/EN/For/ThA/phoenix/index.html} code (a one-dimensional line-by-line code with spherical geometry and full angular integration of specific intensities with a 0.5$^{\circ}$ precision), for composition and thermal profiles typical of the Earth. The version of Phoenix we have used was specifically developed for Earth-like atmospheres \citep{Paillet06}. We find the maximum error on the disk-integrated spectrum to be about 5\% for a monochromatic flux. At most wavelengths, and on the wavelength-integrated flux, the error is smaller than a few percents.

\section{Time series analysis}
\begin{figure*}[th!]
\begin{center}
\includegraphics[width=0.9\textwidth]{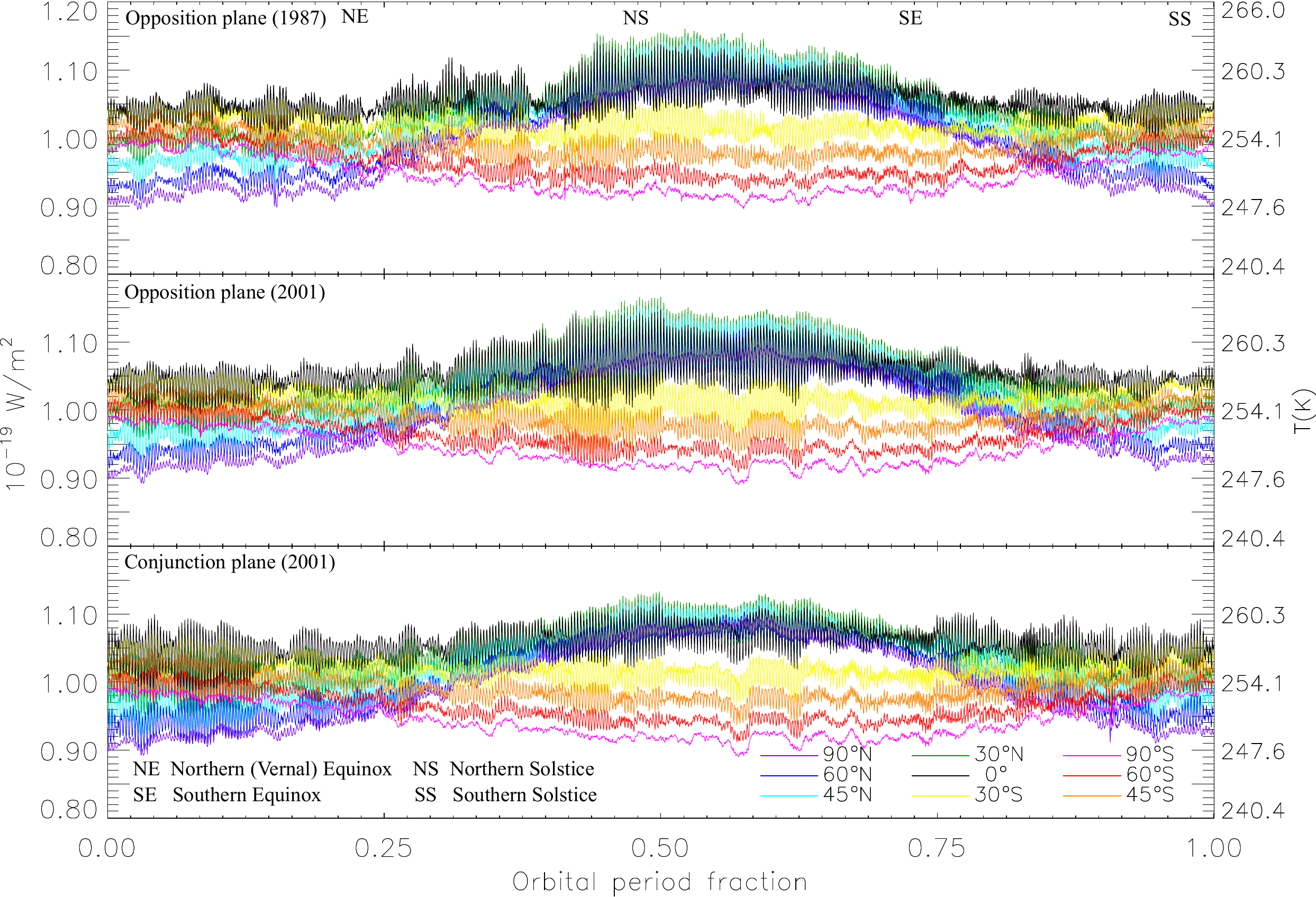}
\caption{Time series of the mid-infrared emission flux for the two years of 1987 (top) and 2001 (middle and bottom). 
The sub-observer's point is represented by the latitudes $90\degr$N, $60\degr$N, 
$45\degr$N, $30\degr$N, $0\degr$, $30\degr$S, $45\degr$S, $60\degr$S, $90\degr$S,
and $0\degr$ longitude at the initial time (January 1, 0:00 UT) and the direction planes of opposition (O) (top and middle) and conjunction (C) (bottom). \label{fig3}}
\par\vspace{5mm}
\includegraphics[width=0.75\textwidth]{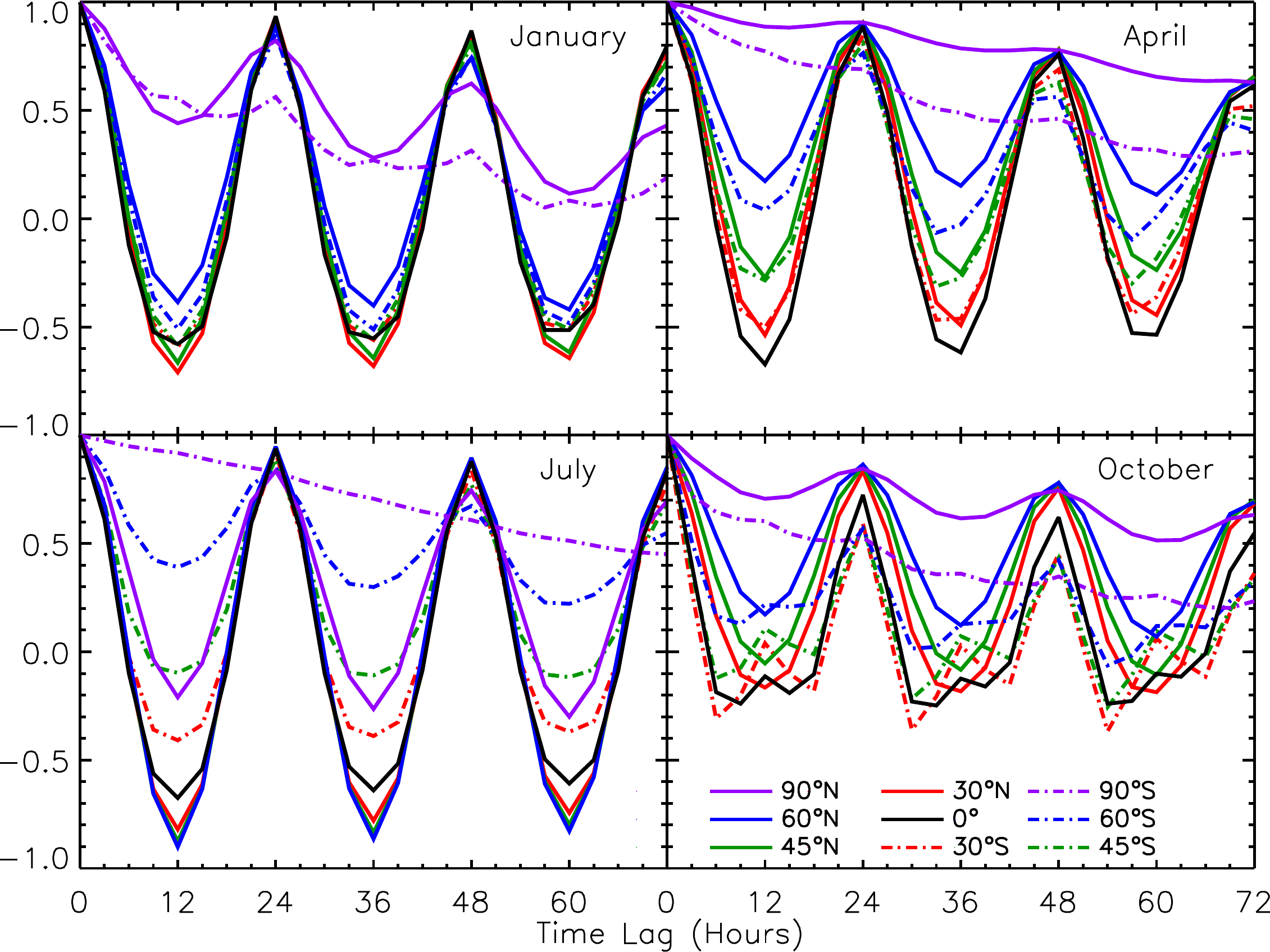}
\caption{Autocorrelation functions of the mid-infrared emission flux from Earth. 
2001 January, April, July, and October for the latitudes $90\degr$N, $60\degr$N,
$45\degr$N, $30\degr$N, $0\degr$, $30\degr$S, $45\degr$S, $60\degr$S and $90\degr$S. \label{fig4}}
\end{center}
\end{figure*}

Two examples of the annual time series of Earth's outgoing mid-infrared radiation are plotted in Figure~\ref{fig3}. The high frequency variability corresponds to the emitted mid-infrared flux due to Earth's rotation super-imposed to the seasonal variation during the year. It is readily observable from the figure that the amplitude of the rotational variability is larger for an observer in the equatorial plane (black) and decreases toward more poleward views (violet, magenta). On the contrary, because the obliquity of the Earth is about $23.44\degr$ and then the variation of the annual insolation is larger at higher latitudes, the amplitude of the seasonal variability increases for the polar geometries and decreases towards the equatorial view. 
As it is expected, the seasonal variation of the northern latitude time series is opposite to the southern ones due to the seasonal cycle of solar insolation. For observers over the Northern Hemisphere, the time series reach a maximum during the boreal summer near the beginning of August (day 211). The equatorial view follows the same pattern, showing a ``northern-biased" behavior of the planet. This effect is due to the existence of large landmasses in the Northern Hemisphere that emit more infrared radiation than the oceans and make the northern summer hotter (Section 3.2). 
Table~\ref{tab1} shows the amplitude values of the seasonal variability of the globally integrated mid-infrared flux for the years of 1987 and 2001 and for five viewing geometries: $0\degr$ (Equator), $45\degr$N, $45\degr$S, $90\degr$N (North Pole) and $90\degr$S (South Pole). The amplitude is given in percentage change over the mean annual value. In order to conduct a further seasonal study of the mid-infrared Earth emission, four representative months (January, April, July and October) were selected per year. For the sake of simplicity, we present in this paper the results of the seasonal study of 2001 as the rest of the data set give similar results. The values given in the table are the values for the observers placed in the opposition plane and in the conjunction plane (in parenthesis), as they have the same view of the planet at different local hours. We can see that the orbital amplitude variation of a northern latitude is twice the value of the southern equivalent latitude. The seasonal variability dominates the signal in most latitudes except for the equator, where seasonal and rotational variability are similar. It is notable that whereas the mean temperature varies with the season, the rotational variability depends strongly on the local hour (Section 3.2). Mid-latitude and equatorial viewing geometries show a pronounced and synchronized rotational variability (up to an 8\% change), whereas the polar light curves present almost no variations.

\begin{deluxetable}{l  ll | llll}
\tablecaption{Photometric Variability of the Integrated Earth Mid-infrared Flux (Percentage Over the Mean Value) } 
\tablehead{ \multicolumn{7}{c}{Photometric Variability (\%)}}
\tablewidth{0pt}
\startdata
{\multirow{2}{*}{Viewing Angle}} & \multicolumn{2}{||c|}{Orbital  Amplitude} &   \multicolumn{4}{c}{Rotational Amplitude}\\ \cline{2-7}& \multicolumn{1}{||c}{1987} & \multicolumn{1}{c|}{2001}& Jan&Apr&Jul&Oct \\\hline
$90\degr$N (N. Pole)& \multicolumn{1}{||c}{20}& \multicolumn{1}{c|}{19}& \multicolumn{1}{c}{1}&\multicolumn{1}{c}{1}&\multicolumn{1}{c}{2}&\multicolumn{1}{c}{1} \\
$45\degr$N (Mid-lat)&  \multicolumn{1}{||c}{14} & \multicolumn{1}{c|}{14}&\multicolumn{1}{c}{4(6)}&\multicolumn{1}{c}{4(4)}&\multicolumn{1}{c}{8(5)}&\multicolumn{1}{c}{5(5)} \\
$0\degr$ (Equator)&  \multicolumn{1}{||c}{6} & \multicolumn{1}{c|}{5}&\multicolumn{1}{c}{5(8)}&\multicolumn{1}{c}{5(4)}&\multicolumn{1}{c}{8(5)}&\multicolumn{1}{c}{4(5)} \\
$45\degr$S (Mid-lat)&  \multicolumn{1}{||c}{7} & \multicolumn{1}{c|}{7}&\multicolumn{1}{c}{3(5)}&\multicolumn{1}{c}{3(2)}&\multicolumn{1}{c}{3(2)}&\multicolumn{1}{c}{2(3)} \\
$90\degr$S (S. Pole)& \multicolumn{1}{||c}{11} & \multicolumn{1}{c|}{10}&\multicolumn{1}{c}{1}&\multicolumn{1}{c}{1}&\multicolumn{1}{c}{1}&\multicolumn{1}{c}{1}\enddata
\tablecomments{Mean amplitude values of the orbital (seasonal) variability for the years of 1987 and 2001 and of the rotational variability over the months of 2001 January, April, July and October for (O) and (C) (in parenthesis) observers situated at different latitudes. Each value is calculated as the percentage of the average variation over the mean value. Data analysis for different years/months retrieves similar results.}
\label{tab1}
\end{deluxetable}

\begin{figure*}[th!]
\begin{center}
\includegraphics[width=0.8\textwidth]{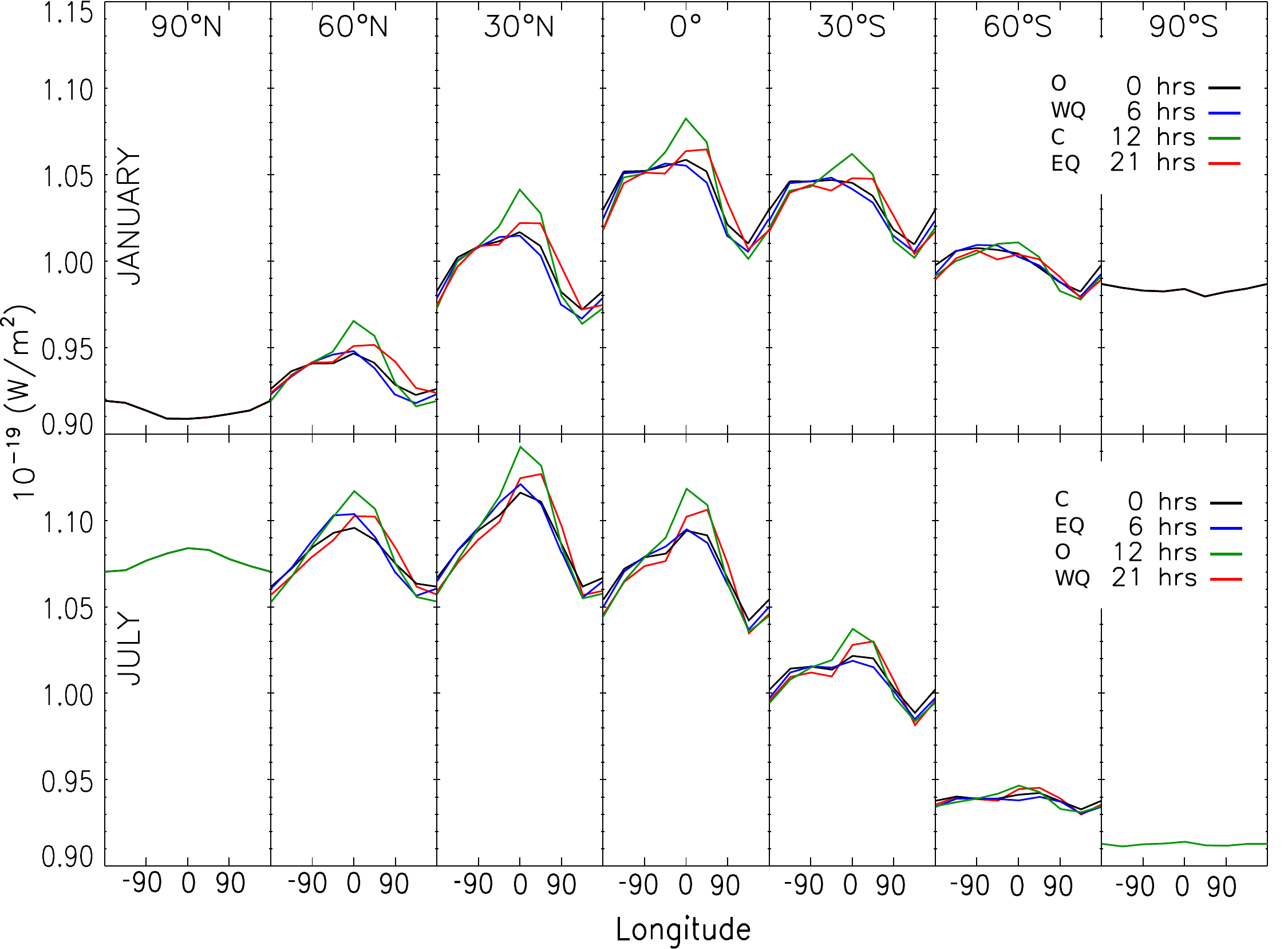}
\caption{Rotational light curves of the mid-infrared radiation emitted from the Earth. 2001 January (top row) and July (bottom row) for the latitudes $90\degr$N, $60\degr$N, $30\degr$N, $0\degr$, $30\degr$S, $60\degr$S, and $90\degr$S latitudes, the colors correspond to local hours, 0 hr (black), 6 hr (blue), 12 hr (green) and 18 hr (red) and the direction planes of opposition (O), conjunction (C), western quadrature (WQ) and eastern quadrature (EQ).\label{fig5}}
\end{center}
\end{figure*}

\subsection{Periodicities}
The analysis of the cross-correlation of the time series with itself or autocorrelation, shown in Figure~\ref{fig4}, is a technique that allows to determine the rotation period of the planet and the lifetime of the cloud structures.  The autocorrelation $A$ can be computed as:

\begin{equation}\label{eqauto}
A(L)=\frac{\sum_{k=0}^{N-L-1}\left(F(k\Delta t) -\overline{F}\right )\left(F((k+L)\Delta t)-\overline{F} \right )}{\sum_{k=0}^{N-1}\left(F(k\Delta t)-\overline{F} \right )^{2}},
\end{equation}

where $L$ is the time lag in number of 3 hr points, $N$ is the total number of points in the time series, and $\overline{F}$ is the mean flux. The autocorrelation is maximum when values are similar within a time lag distance and it is sampled according to the time resolution. This method was chosen before the Fourier Transform to avoid the harmonics of the rotational period. The duration of the statistically significant peaks in the autocorrelated time series can give us an estimation of the lifetime of the cloud structures, typically of around one week for Earth clouds \citep{Palle2008}. In the outgoing mid-infrared radiation flux, a 24 hr rotation period is clearly shown, a value close to the true rotation period. This rotational signature has two origins. First, some large regions exhibit systematically high or low brightness temperatures. This is the case for Indonesian and Sahara areas, as the former is one of the most humid and cloudy regions on the planet, whereas the latter is warmer and drier than Earth's average. Therefore, the two regions appear as fixed cold and hot features respectively, even on averaged outgoing flux maps, as we can see in Figure~\ref{fig1}(c) and (d) where the TOA emission is averaged over a month and in Figure~\ref{fig1}(e) and (f) where it is averaged over a year. A smaller effect comes from the diurnal cycle (the change of brightness temperature between day and night in a region), which is negligible in most locations, because of humidity, clouds, or ocean thermal inertia, but noticeable in some dry continental areas as it is shown in Figure~\ref{fig5}.

\begin{figure*}[th!]
\begin{center}
\includegraphics[width=0.8\textwidth]{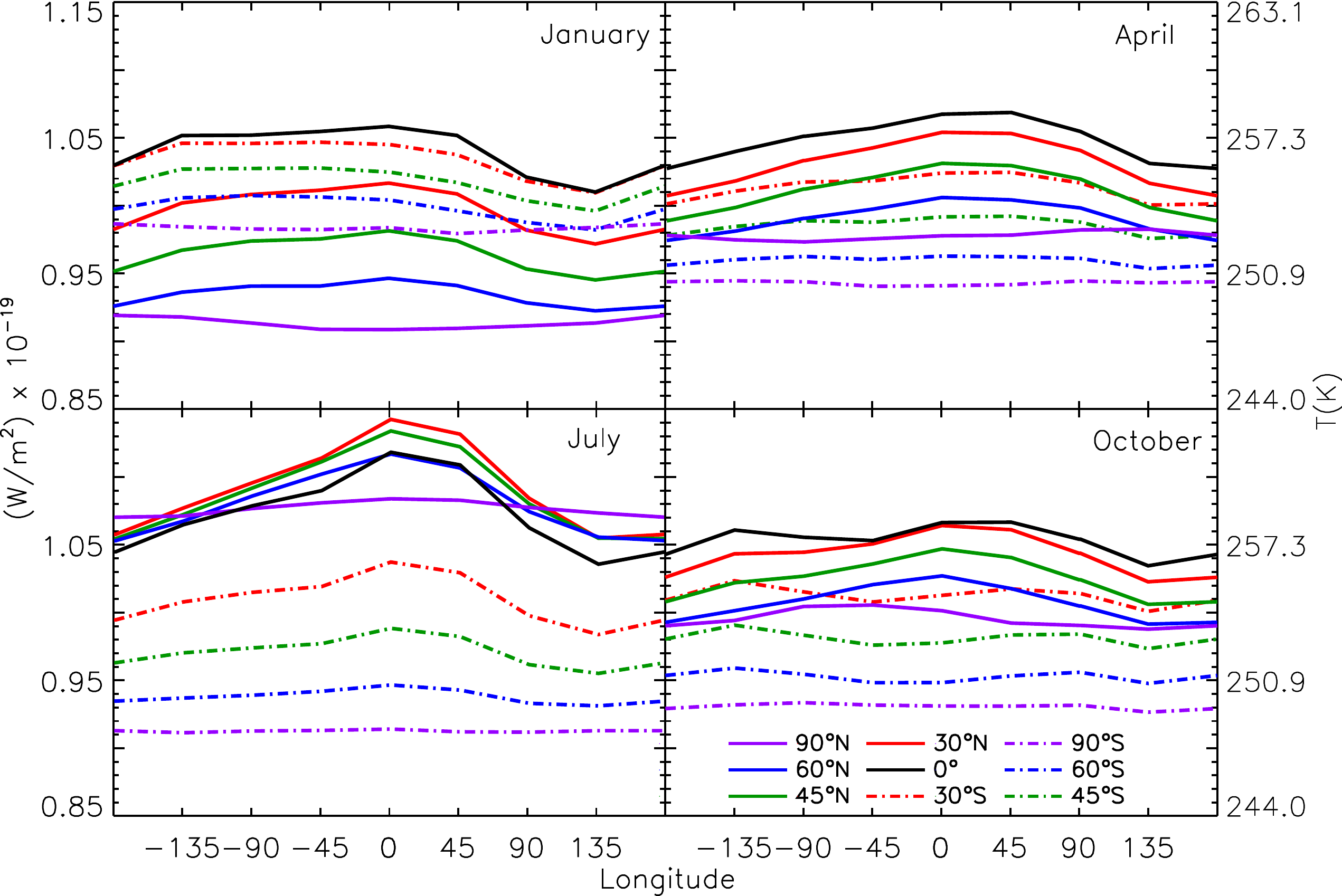}
\caption{Rotation light curves of the mid-infrared radiation emitted from the Earth. 
For the months of 2001 January, April, July and October, $0\degr$ longitude and
 latitudes: $90\degr$N, $60\degr$N, $45\degr$N, $30\degr$N, $0\degr$, $30\degr$S, 
 $45\degr$S, $60\degr$S, and $90\degr$S.\label{fig6}}
\end{center}
\end{figure*}
In fact, it is the diurnal cycle of the dry lands, which makes possible the detection of the rotation period for the case of the North-polar view during the more stable seasons (winter and summer), as it is shown in Section 3.2. Although it is always the same fraction of the planet (the whole northern hemisphere) in the field of view, the change in temperature along the day in the dry continental areas causes the rotational modulation as it is shown in Figure~\ref{fig4} (violet solid line). However, an observer looking at the South Pole does not detect a significant rotational variability, Figure~\ref{fig4} (violet dash-dotted line), not even for the austral summer when the effect of clouds would be minimized. In the Southern Hemisphere, the distribution of land is largely dominated by oceans whose high thermal inertia make diurnal temperature variability negligible.

\subsection{Average Rotation Light Curves}
Once the rotation period is identified, the observer can produce a typical rotation light curve by folding the time series obtained during weeks or months over the rotation period. This process averages out random cloud variability. It is clearly seen on the maps of Figures~\ref{fig1}(c)--(f), where the clouds disappear for longer average times whereas the strongest features mentioned on the previous section prevail. Then the observer can plot this average rotation light curve as a function of an arbitrary longitude (in our case the longitude that we have chosen is the conventional geographic longitude for commodity). The shape of the light curve can then reveal brighter/fainter areas distributed in longitude. For instance, observers over a latitude of $30\degr$N would note that the brightest and faintest point of the light curve occur when the longitude $0\degr$ (Sahara) or $135\degr$ (Indonesia) are respectively centered on his view. The results are represented in Figures~\ref{fig5} and  ~\ref{fig6}. 

Figure~\ref{fig5} represents the rotational variability with local hour for several latitudes (in column) and at the months of January (top row) and July (bottom row). Each chart corresponds to four observers placed at the same latitude referred and at the four initial longitudes O, C, WQ, EQ, previously defined. Each of these four planes correspond to a certain local hour that changes along the orbital period. With this information, we can compare the temperature evolution for a given planetary region along the day (diurnal variability). The maxima show the largest variation temperature along the day whereas the minima hardly vary. For the cases where the observers are placed over the poles, the planetary view does not change with time so the local hour of the graph is taken just as reference of the observation time. For the North Polar case and for both summer and winter seasons, the minima occurs when it is 0 UT (0 hr at $0\degr$ longitude or 12 hr at $180\degr$ longitude) and the maxima when it is 12 UT (0 hr at $180\degr$ longitude or 12 hr at $0\degr$ longitude), which illustrates the diurnal cycle effect previously mentioned. These hours coincide respectively to midnight and midday in the Sahara desert. The greatest influence of the area is noticed at $30\degr$N latitude when the Sahara is in the center of the planetary disk. In summer or winter the diurnal variability reaches the 2\%, although the main contribution to the flux is that of the solar insolation along the year. 

Figure~\ref{fig6} represents the rotational variability for observers placed at $0\degr$ longitude and different latitudes.  It is important to note the temperature evolution with time for each geometry, implying a seasonal behavior. The equator shows a warm stable temperature during the year, whereas the summers of each hemisphere differ being the northern summer hotter, as it is shown in Figure~\ref{fig5}. The maximum and minimum regions do not change with the seasons, except for April and October when the presence of the clouds can mask the signal. 

\begin{figure*}[th!]
\begin{center}
\includegraphics[width=0.8\textwidth]{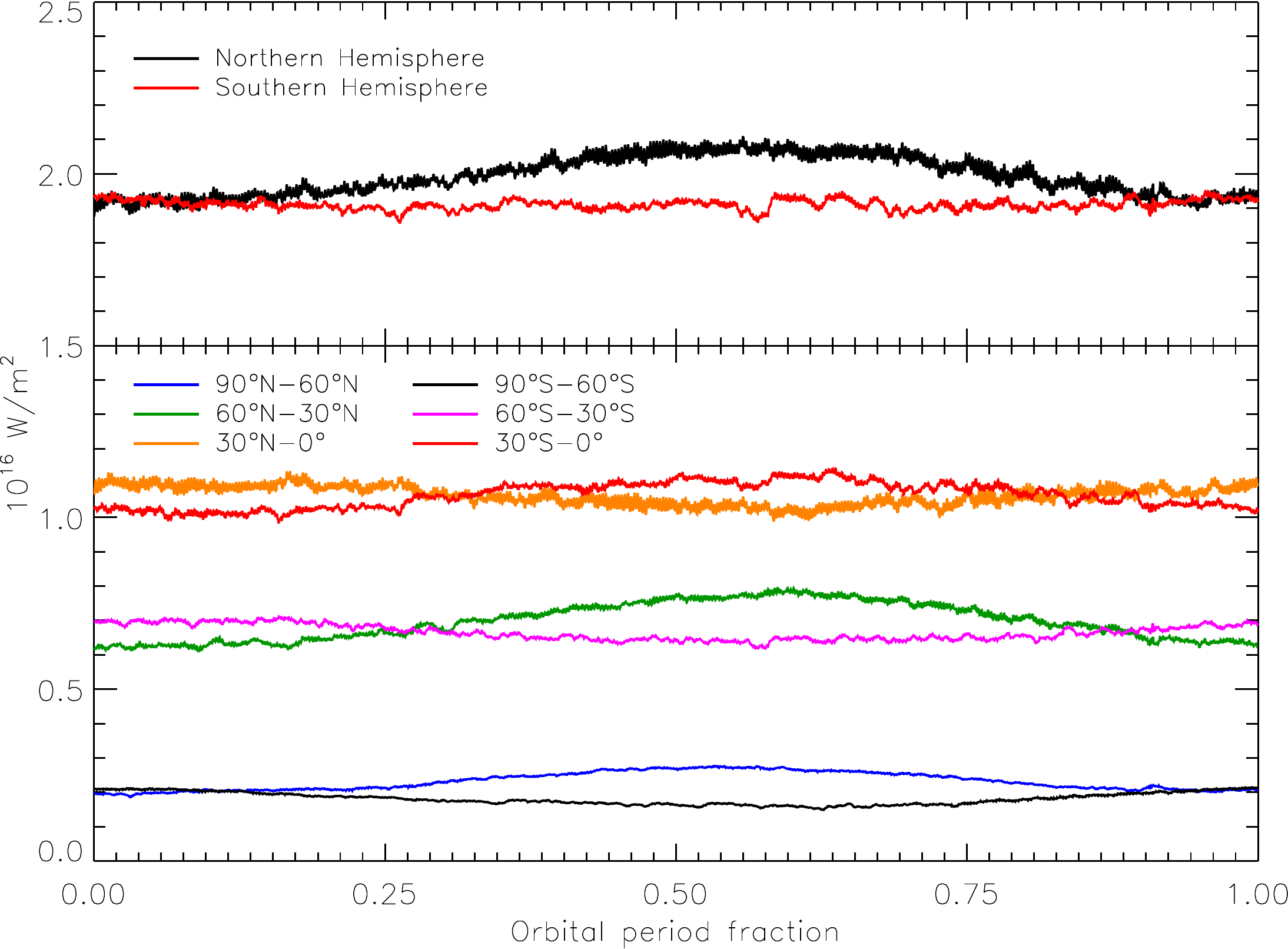}
\caption{TOA--LW infrared emission flux of the Northern and Southern hemispheres. Total emission (top) of the Northern Hemisphere (black) and Southern Hemisphere (red). Latitude bands (bottom) of $90\degr$N--$60\degr$N (blue), $60\degr$N--$30\degr$N (green), $30\degr$N--$0\degr$ (orange), $90\degr$S--$60\degr$S (black), $60\degr$S--$30\degr$S (magenta) and  $30\degr$S--$0\degr$ (red).\label{fig7}}
\end{center}
\end{figure*}

\begin{figure*}[th!]
\begin{center}
\includegraphics[width=0.81\textwidth]{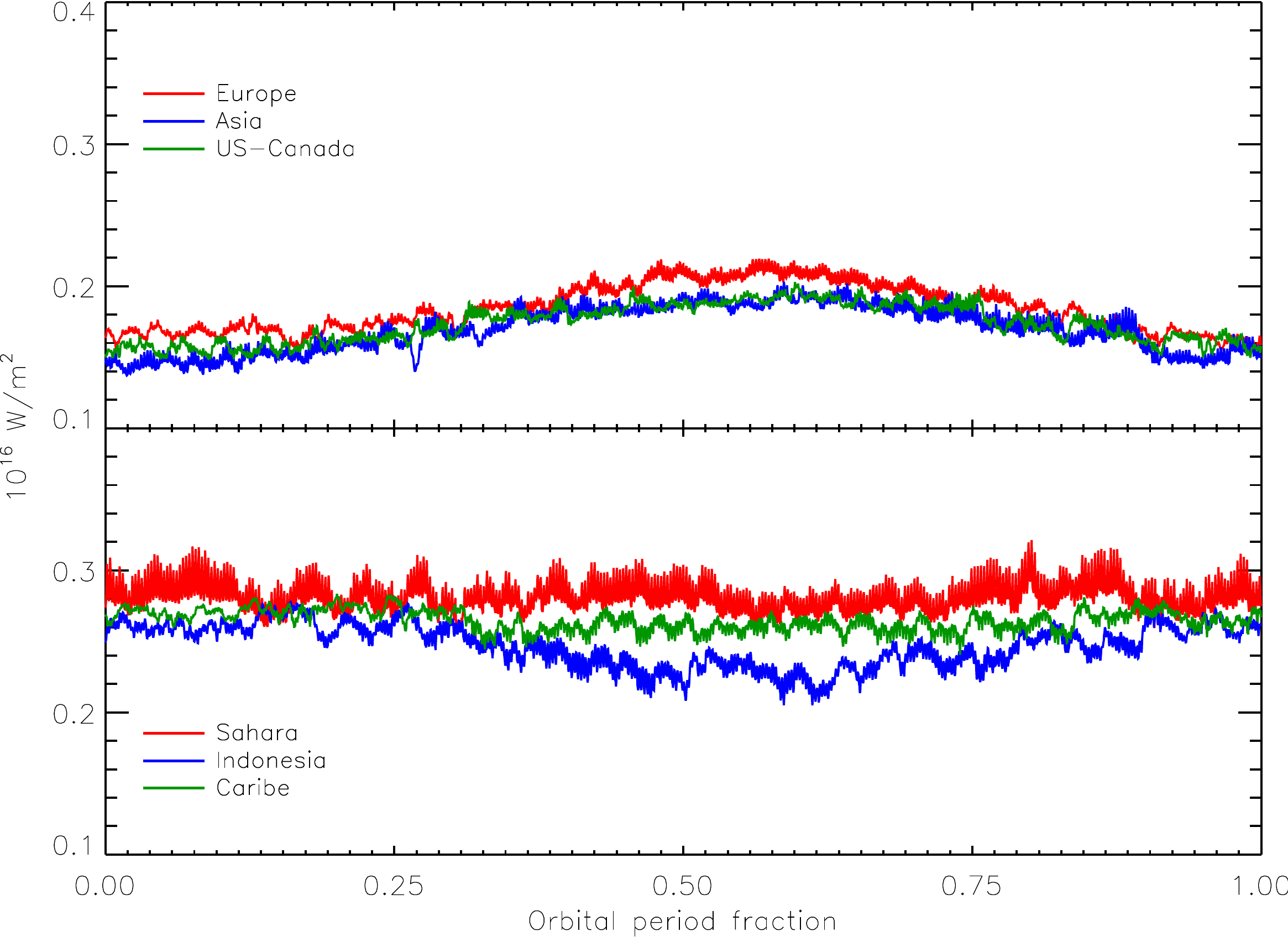}
\caption{TOA--LW infrared emission flux of some continental regions in the Northern Hemisphere. In the $60\degr$N--$30\degr$N latitude band (top): Europe (red), Asia (blue), and US-Canada (green). In the $30\degr$N--$0\degr$ latitude band (bottom), Sahara-Arabian (red), Indonesian (blue), and Caribbean-Mexico area (green).\label{fig8}}
\par\vspace{5mm}
\includegraphics[width=0.81\textwidth]{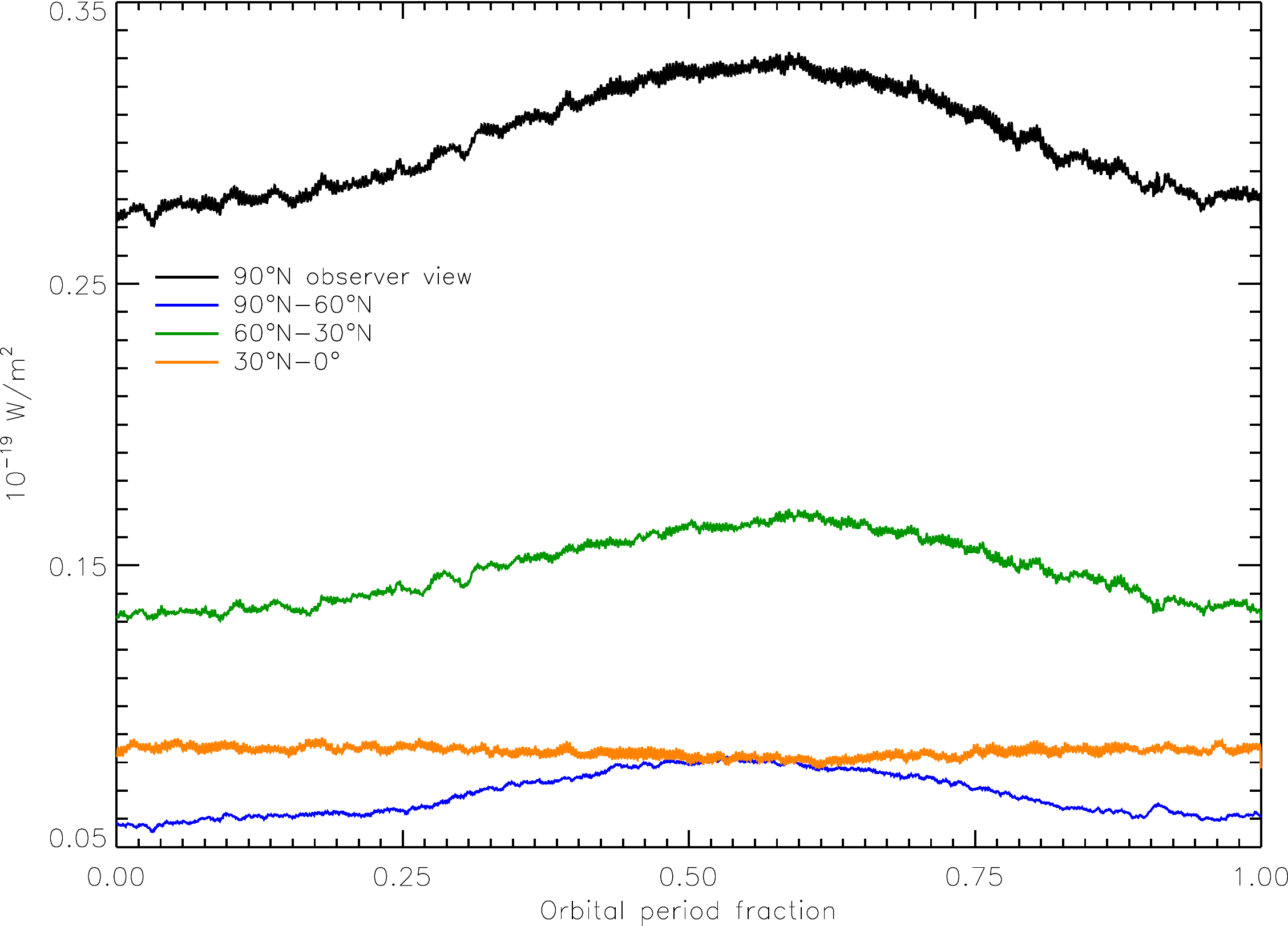}
\caption{Time series for a North Polar observer (black) and the contribution of each latitude band to the signal. $90\degr$N--$60\degr$N latitude band (blue), 
$60\degr$N--$30\degr$N (green) and $30\degr$N--$0\degr$ (orange).\label{fig9}}
\end{center}
\end{figure*}

 In order to check the possible source of the diurnal variation on the Northern Hemisphere, we have made a further analysis of the emitted flux by geographic latitude bands. Figure~\ref{fig7} shows the mid-infrared emission of the Earth along one orbital period. Each hemisphere is divided in three latitude bands of flux: $90\degr$N--$60\degr$N (blue), $60\degr$N--$30\degr$N (green), $30\degr$N--$0\degr$ (orange), $90\degr$S--$60\degr$S (black), $60\degr$S--$30\degr$S (magenta), $30\degr$S--$0\degr$ (red). We can see that the Northern Hemisphere exhibits a bigger seasonal variation than the Southern Hemisphere and this variation is more important for mid-latitudes. As it was previously mentioned, the greater diurnal variability comes from the band $30\degr$N--$0\degr$. Besides, in this band, there is a decrease in flux during summer. This is due to the migration of the Intertropical Convergence Zone to these latitudes, which produces large bands of high humidity and clouds that have lower brightness temperatures.  As expected, the South Pole is colder than the North Pole.
For the identification of the region that produces the biggest variability, we have analyzed the emission of six regions of the planet (Figure~\ref{fig8}). The chosen regions have an area of $30\degr$$\times$$60\degr$ latitude--longitude, then fluxes from the same latitude band can be compared. For the $60\degr$N--$30\degr$N band we have chosen three regions centered in Europe (red), Asia (blue) and US-Canada (green). For the $30\degr$N--$0\degr$ band the regions are centered in the Sahara-Arabian deserts (red), Indonesia (blue), and the Caribbean-Mexico area (green). As expected, the regions in the same latitude band have similar seasonal variabilities but the Sahara region emits more infrared flux and has greater diurnal variability. However, the emission received by the observer changes with the view of the planet. The Sahara desert lies near the limb of the planetary disk facing an observer placed over the North Pole and its influence on the signal is lessened by the perspective. In Figure~\ref{fig9}, we can see that the diurnal variability of the signal has a contribution of both low and mid-latitudes. Finally comparing Figures~\ref{fig5}--\ref{fig7} with the signal received, shown in Figure~\ref{fig3}, we conclude that although there is a small difference between day and night produced by the warm areas, the signal is dominated by the seasonal behavior.

\subsection{Phase Variation}
While Earth's visible flux received by a remote observer is modulated by the changing phase of the planet, the Earth does not present significant phases when the integrated infrared flux is observed: the emission from the nightside contributes nearly as much as the emission from the dayside \citep{Selsis2004}. This is shown in Figure~\ref{fig3} (middle and bottom), which represent opposite observers and then opposite visible phases of the planet. While observer (O) sees the winter midnight, the observer (C) sees the winter noon which shows the same average temperature but bigger rotational variation. Figure~\ref{fig5} shows this difference between 12 hr and 0 hr, however the seasonal variation dominates. Continental surfaces and the boundary layer above them (roughly the first 1500 m of the atmosphere) experience a drop of temperature between the day and night. This diurnal cycle is insignificant above the ocean (due to the high thermal inertia of water), hence over $\sim70\%$ of the Earth surface. Day--night brightness temperature variations affect the outgoing thermal emission only in the 8--12$\mu$m atmospheric window and above dry continents. This happens either over very cold regions which in this case do not contribute much to the global emission, or over deserts which represent a small fraction of the Earth surface. In addition, the diurnal cycle is much less pronounced above the boundary layer, at altitudes where most of the thermal emission is emitted to space. This is the reason why phase-correlated variations of Earth brightness temperature are negligible compared with seasonal changes. Even for the observing latitudes in the $30\degr$N--$0\degr$ range, in which the Sahara diurnal cycle appears in the modulation, the winter midday is colder than the summer midnight as it is seen in Figure~\ref{fig5}.\\ 

\begin{figure*}[th!]
\begin{center}
\includegraphics[width=0.85\textwidth]{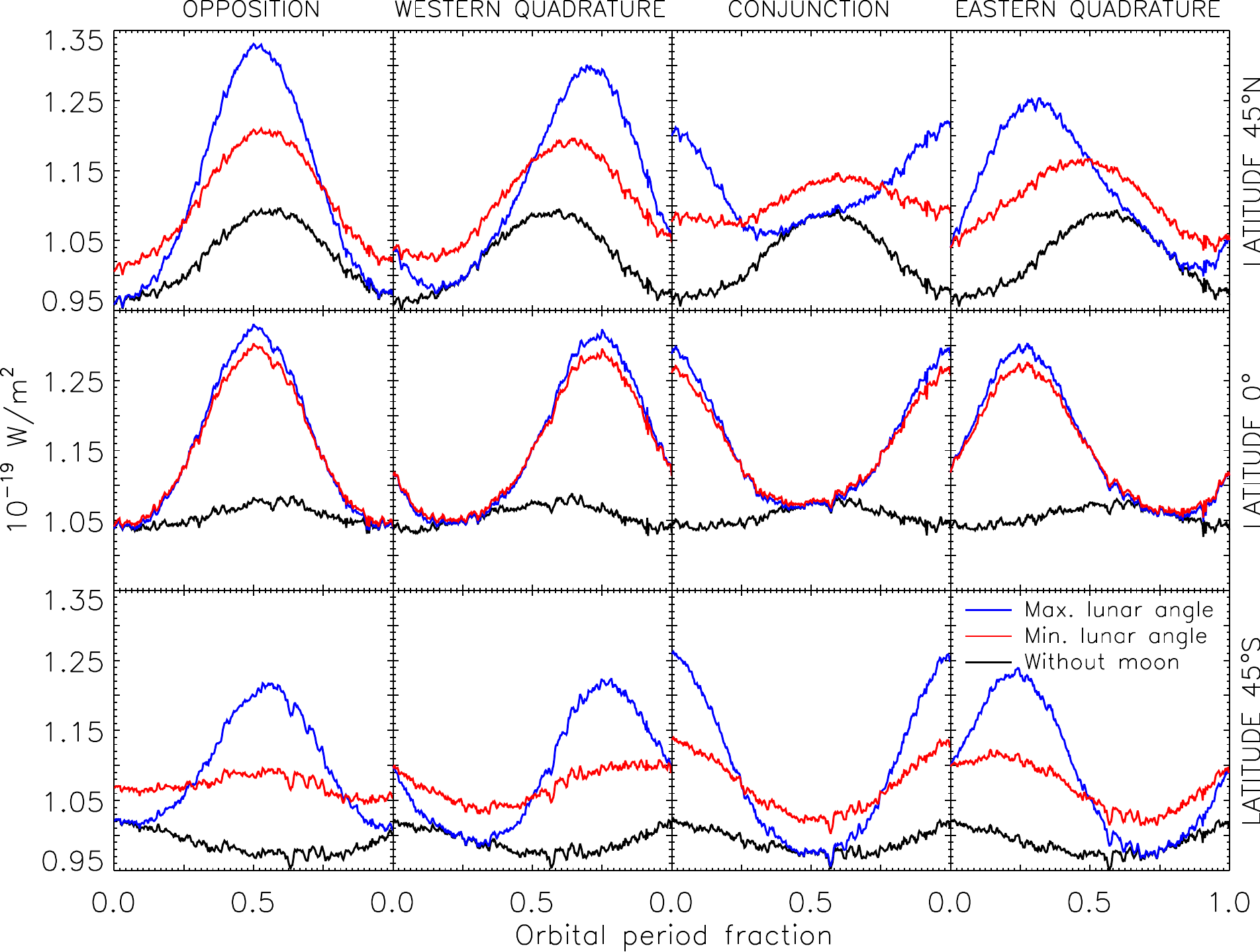}
\caption{Earth--Moon mid-infrared emission light curves for one planetary orbit. 
$45\degr$N (top row),  Equator (second row), $45\degr$S latitudes (bottom row), in columns the signals received by the observer's placed in opposition, western quadrature, conjunction, and eastern quadrature at the initial time. The colors correspond to the Earth (black) and Earth--Moon system, with lowest (red) and highest (blue) inclination angles of the Moon's orbit according to the observer's geometry (the possible orbits are comprehended between the two). \label{fig10}}
\end{center}
\end{figure*}

\subsection{Earth--Moon System Light Curves}

We have modeled the mid-infrared flux of the Moon. Due to a very low surface thermal inertia, the temperature map of the starlit hemisphere of the Moon can be calculated by assuming local radiative equilibrium at the surface \citep{Lawson2007}. When calculating the disk-integrated emission, the contribution of the dark side can be safely neglected due to the high temperature difference. Thus, the flux received depends only on the phase of the Moon as seen by the distant observer. The amplitude of the lunar phase variations depends on the elevation of the observer above the lunar orbit, when we computed the Earth signal as a function of the latitude of the observer. To a given latitude of the observer, it corresponds to a range of possible elevations above the lunar orbit. In Figure~\ref{fig10}, instead of calculating this elevation consistently with the chosen observer geometry, we bracket the orbital light curve of the unresolved Earth--Moon system with the two curves obtained by adding the lunar signal for the two extreme possible elevations. As pointed by \citet{Selsis2004} and \citet{Mosko09}, it presents phase variations dominated by the Moon. We note that the modulation from the satellite becomes negligible for a Moon-like satellite with $20\%$ of the Moon radius, a $\sim 5\%$ of the radius of the planet. 

The two main annual variations that modulate the IR emission from the point-like Earth--Moon system are due to the seasons of the Earth and the phases of the Moon. These modulations present a phase shift that depends on the observer geometry. For some geometries, these two modulations are coincidental. This happens if the maximum of the lunar phase corresponds to Earth's annual emission maximum, which is for instance the case for an observer looking at northern latitudes that sees the Sahara at noon in July (see Figure~\ref{fig10}, the first two panels of the left column). This particular observer will see only phase-correlated variations and may attribute this variability to a day--night temperature difference and conclude that the planet has less ocean coverage and a thinner atmosphere. With such particular geometry, seasonal variations could also be mistakenly attributed to the phases in the absence of a moon, unless the lunar origin of the modulation is identified using spectroscopy \citep{Robinson2011b}. 

\section{Conclusions}
In this paper, we have constructed a 3 hr resolution model of the integrated mid-infrared emission of the Earth over 20 years in the direction of a remote observer randomly located. 

The seasonal modulation dominates the variation of the signal. As expected, it is larger for the polar views because the planetary obliquity causes a bigger annual insolation change for these latitudes. For equatorial views, the seasonal maximum occurs during the summer of the Northern Hemisphere, as the latter contains large continental masses whereas the Southern Hemisphere is dominated by the oceans. 

The rotational variability is detectable because of the uneven distribution of oceans and continents with geographical longitude. The daily maximum of the mid-infrared flux is shown when dry large masses of land, such as the Sahara desert, are in the observer's field of view. The daily minimum appears when cloudy humid regions such as the Indonesian area is visible, as iced big zone are confined to the poles. In the polar views, the distribution of land does not change with time but the diurnal temperature variation of large continental areas affects the signal, allowing the detection of the rotational period in the North Polar case. We find that the rotational variations have an amplitude of several percent, which is comparable to that of the seasonal variations for some latitudes. It is important to remark the strong influence of the weather patterns, humidity and clouds are sometimes able to mask the 24 hr rotation period of the signal for several days at a time. However, this effect can be solved by time folding.

 It is important to point out that the Earth does not exhibit a significant modulation associated with phase variation (phase curve). This is because the integrated thermal emission does not generally probe the boundary layer (first km of the atmosphere) where the diurnal cycle takes place. If unresolved, the Earth--Moon system would however present a phase variation of Lunar origin. A satellite of the size of the Moon would introduce a strong phase variability that would completely dominate over the planet's signal. This effect adds high complexity to its interpretation by photometry. 

At the light of these results it seems that future infrared photometric observations of terrestrial planets can be useful in order to characterize their atmospheric and surface features. If the planet is not completely covered by clouds, as Venus is, the presence of strong surface inhomogeneities (continents) can be extracted from the daily variations. The seasonal cycle can also give estimates of the planet effective temperature, the variability in the obliquity of its orbit, and the distribution of land at larger scale. A further study with a Global Circulation Model combining Earth's emitted flux is ongoing. 

\acknowledgments

Research by E. Pall\'e acknowledges support from the Spanish MICIIN, grant CGL2009-10641. F. Selsis acknowledges support from the European Research Council (starting grant 209622: E$_3$ARTHs).

\end{document}